\documentstyle[epsf,aps,prl,multicol]{revtex}

\begin{document}
\draft
\title{An efficient scheme for the deterministic maximal entanglement 
of N trapped ions}
\author{J.~Steinbach and C.C.~Gerry\footnote{Permanent address: 
Lehman College, The City University of New York, 250 Bedford Park
Boulevard West, Bronx, NY 10468-1589, U.S.A.}}
\address{Optics Section, Blackett Laboratory, Imperial College,
London SW7 2BZ, United Kingdom}
\date{November 9, 1998}
\maketitle
\begin{abstract}
We propose a method for generating maximally entangled states of $N$
two-level trapped ions. The method is deterministic and independent of
the number of ions in the trap. It involves a controlled-NOT acting
simultaneously on all the ions through a dispersive interaction. We
explore the potential application of our scheme for high precision
frequency standards.
\end{abstract}

\begin{multicols}{2}

The entanglement of quantum states of two or more particles, aside
from being of intrinsic interest, is of great practical importance in
the fields of quantum cryptography and quantum computation
\cite{quantcomp}. One other area where entangled quantum states
may have a significant impact is that of the improvement of frequency
standards \cite{freqstd,wineland98,bollinger96}. Advances in cooling
and trapping of ions have given rise to new techniques in high
precision spectroscopy which may yield frequency standards with
accuracies of the order of one part in $10^{14}$--$10^{18}$
\cite{spec}. Key to the improvement of frequency standards beyond the
shot-noise limit \cite{freqstd} is the establishment of an entangled
state of a collection of $N$ two-level atoms. Initial theoretical
investigations examined the use of squeezed spin states
\cite{freqstd,wineland98}. We concentrate here on a {\it maximally\/}
entangled $N$-particle state, having the form
\cite{bollinger96}
\begin{equation}
 |\Psi_M\rangle = \frac{1}{\sqrt{2}} \left\{|e_1,e_2,...e_N\rangle 
	+ \,e^{i\phi} |g_1,g_2,...g_N\rangle \right\}\,,
 \label{10}
\end{equation}
where $|e_{j}\rangle$ and $|g_{j}\rangle$ denote the excited and
ground states of the $j$th particle respectively. Using the Dicke
angular momentum states \cite{dicke54} this state can be written as
\begin{equation}
 |\Psi_M\rangle = \frac{1}{\sqrt{2}} \left\{ |J,J\rangle + \,e^{i\phi}
 |J,-J\rangle \right\}\,,
 \label{20}
\end{equation}
where $J = N/2.$ The above state is an $N$-particle version of the
Greenberger-Horne-Zeilinger state \cite{GHZ} and has been shown to
display extreme quantum entanglement \cite{mermin90}. It may also be
considered a special case of the atomic Schr\"odinger cat states
\cite{atomcat}. Recently, Turchette {\it et~al.} have reported the
generation of a non maximally entangled two-particle state using a
deterministic method in an ion trap experiment \cite{turchette98}.

The maximally entangled state given in Eq.\,(\ref{20}) may be used in
high precision spectroscopy to measure the transition frequency
$\omega_0 = (E_e - E_g)/\hbar,$ where $E_e$ and $E_g$ are the
respective energies of the electronic excited and ground states
\cite{bollinger96}. In contrast to measurements with uncorrelated
atoms, which yield an uncertainty in the frequency that depends on
$N^{-1/2},$ the state $|\Psi_M\rangle$ allows one to measure the
transition frequency to an uncertainty of $N^{-1}.$ Huelga {\it
et~al.} \cite{huelga97} have described how in principle a standard
Ramsey spectroscopy scheme \cite{ramsey63} can be modified to achieve
this limit. The key to this is a controlled-NOT (C-NOT) operation
right after the initial and before the final Ramsey pulse, where the
electronic state of the ion which is manipulated through the Ramsey
pulses acts as a control to flip the electronic state of all the other
ions.

As Huelga {\it et~al.} \cite{huelga97} have shown, in the presence of
decoherence the standard Ramsey spectroscopy measurements on
uncorrelated atoms and measurements on the maximally entangled states
yield the same resolution. Of course, if decoherence is not present or
if the measurements can be done in a time short compared to the
decoherence time, the maximally entangled states will yield higher
resolution frequency measurements and thus it is of interest to find
efficient mechanisms for their generation.

In view of recent experimental progress \cite{turchette98}, perhaps
the most promising physical system for the generation of the type of
maximally entangled state given in Eq.\,(\ref{20}) is a string of
laser-cooled ions in a linear rf trap. Previously, Cirac and Zoller
\cite{cirac95} have proposed a method which sequentially performs $N$
C-NOT operations between the internal states of pairs of ions in such
a string. This approach requires individually addressing all the ions
with a well focused laser beam. Bollinger {\it et~al.}
\cite{bollinger96} have proposed an alternative method that does not
require interacting with the ions individually. It {\it does} require
the use of three vibrational modes and the generation of linear
couplings between pairs of those modes as electronic transitions are
driven for all the ions simultaneously \cite{steinbach97}. Both in the
scheme by Cirac and Zoller \cite{cirac95} and the method proposed by
Bollinger {\it et~al.} \cite{bollinger96} the number of steps or laser
pulses required to generate the maximally entangled state
$|\Psi_M\rangle$ is proportional to the number of ions. Only recently,
Wineland {\em et~al.} \cite{wineland98} have proposed for the first
time a sequence of operations which accomplishes this with a fixed
number of steps.

In this paper, we present a method of generating states of the form of
Eq.\,(\ref{20}), which is independent of the number of ions, and
in contrast to most experiments that generate entangled states, is
deterministic \cite{deterministic}. While our scheme is related to the
proposal in \cite{wineland98} through the sequence of operations which
lead to the maximally entangled state, it is significantly different
in that (i) it does not rely on a specific value of the Lamb-Dicke
parameter, (ii) it operates only on two electronic levels of the
trapped ions and (iii) it points out a C-NOT operation which can
operate on multiple ions. Our scheme requires that {\it one\/} of the
ions be addressed individually for certain manipulations and we assume
here that this is done with a well focused laser
\cite{hughes97}. We further assume that all the ions can be
addressed simultaneously with a laser beam of sufficiently broad waist
\cite{turchette98}. The maximally entangled state $|\Psi_M\rangle$ is
generated through a sequence of five laser pulses, starting from an
initial state $|J,-J\rangle|0\rangle,$ where all the ions are in their
ground state and the collective motion has been cooled to the ground
state \cite{king98}. At the heart of our preparation scheme lies a
C-NOT operation between the collective vibrational motion and the
internal states of {\it all\/} the ions. This is generated by a type
of dispersive interaction between those degrees of
freedom. Previously, this kind of coupling has been discussed in
connection with the generation of the vibrational Schr\"odinger cat
states for a single trapped ion \cite{gerry97}, and as a degenerate
Raman-coupled model in the context of cavity quantum electrodynamics
\cite{knight86}.

Before going through the preparation scheme step by step, we consider
the four types of pulses involved. We require pulses at the two
frequencies $\omega_0$ and $\omega_0 - \nu_x,$ where $\nu_x$ is the
frequency of the ions' collective harmonic motion along the trap axis
which we take to be the $x$ axis \cite{harmonic}. We assume that all
pulses are performed with laser beams that are derived from the same
source, so that at time $t=0,$ all electric fields that excite the
system during the preparation scheme are in phase. To derive the
transformations caused by the various pulses we generically assume the
respective laser beams to be turned on from time $t = t_0,$ to $t =
t_0 + t_p,$ and we give the corresponding unitary transformations
$U(t_0,t_p),$ in a rotating frame where $|\Psi_{R} (t)\rangle =
\exp{[i \hat{H}_{R} t/\hbar]} |\Psi (t)\rangle$, $\hat{H}_{R} = \hbar
\omega_0 \sum_{i=1}^N |e_{i}\rangle\langle e_{i}|$, and $|\Psi
(t)\rangle$ is the state of the system in the Schr\"odinger
picture. In our analysis of the dynamics generated by the laser
excitation we explicitely consider only the collective motion along
the trap axis, characterized in terms of the vibrational energy
eigenstates $|n\rangle.$ The motional state perpendicular to the trap
axis remains in the ground state throughout the preparation scheme
since we do not excite sidebands of the motion along those directions
and in a linear trap $\nu_x \ll \nu_y,\nu_z.$ We further assume both
the Lamb-Dicke limit \cite{LDL} and the low excitation regime
\cite{cirac96}. The first condition allows us to expand the
Hamiltonian describing the interaction of the ions with the laser
light in terms of the Lamb-Dicke parameter $\eta_x = k_x \Delta x_0$,
where $k_x$ is the projection of the laser wave vector onto the trap
axis and $\Delta x_0 = (\hbar / 2 M \nu_x)^{1/2},$ is the width of the
motional ground state for a single ion along that axis, $M$ being the
mass of a single ion. In the low excitation regime we retain only
resonant transitions in the analysis of the excitation.

We first consider the action of the pulses involving only one of the
ions. The first of these is a resonant $\pi/2$-pulse where a laser
beam of frequency $\omega_0$, propagating perpendicular to the trap
axis excites, say, the $N$th ion. This generates the transformation
\begin{eqnarray}
 U(t_0,t_{\pi/2}) &=& \sum_{n=0}^\infty 
	\frac{1}{\sqrt{2}}\,e^{-i n \nu_x t_{\pi/2}}\, 
	|n\rangle\langle n| \nonumber \\[1ex]
 &&	\otimes \left\{|e_{N}\rangle\langle e_{N}| + 
	|g_{N}\rangle\langle g_{N}|\right. 
	\nonumber \\[1ex]
 &&	- \left. |g_{N}\rangle\langle e_{N}| + 
	|e_{N}\rangle\langle g_{N}|\right\}\,,
 \label{30}
\end{eqnarray}
where the pulse duration $t_{\pi/2}=\pi/2\Omega,$ and $\Omega$ denotes
the Rabi frequency for that pulse. The phase factor is due to the free
evolution of the vibrational degree of freedom during the pulse.

The second type of pulse required is a $\pi$-pulse generated by a
laser of frequency $\omega_0-\nu_x,$ and where the wave vector has a
component $k_x$ along the trap axis. This generates a
$|g_{N}\rangle|n+1\rangle \Leftrightarrow |e_{N}\rangle|n\rangle$
transition of the Jaynes-Cummings type, where the Rabi frequency
$\Omega_{\it JC}^{(n)} \propto \eta_x \sqrt{n+1}/\sqrt{N}.$ For a
given $|g_{N}\rangle|n+1\rangle \Leftrightarrow
|e_{N}\rangle|n\rangle$ transition a pulse of duration $t^{(n)}_{\it
JC, \pi} = \pi/\Omega_{\it JC}^{(n)},$ causes the transformation
\begin{eqnarray}
 && U_{\it JC}^{(n)}(t_{0},t^{(n)}_{\it JC, \pi}) = 
 \nonumber \\[1ex]
 && i\,e^{-i \nu_x\left[t_0 + (n+1) t^{(n)}_{\it JC, \pi}\right]}
 |n+1\rangle\langle n|\otimes|g_{N}\rangle\langle e_{N}|
 \nonumber \\[1ex] 
 && + \ i\,e^{i\nu_x\left[t_0 - n t^{(n)}_{\it JC, \pi}\right]}
 |n\rangle\langle n+1|\otimes |e_{N}\rangle\langle g_{N}|\,,
 \label{40}
\end{eqnarray}
where the phase factors are due to the free evolution of the
vibrational degree of freedom during the pulse and the fact that we
have assumed all electric fields to be in phase at time $t=0.$ We note
that the state $|g_{N}\rangle|0\rangle$ is not coupled by the
Jaynes-Cummings pulse.

The third type of pulse which we shall refer to as a dispersive
$\pi$-pulse is generated by two laser beams of frequency $\omega_0$
\cite{gerry97}. More specifically, the first beam is propagating
perpendicular to the trap axis and the second one has a wave vector
component $k_x$ along that axis.  While not exciting any vibrational
sidebands the pulse exploits the dependence of the generated Rabi
oscillations between the states $|g_{N}\rangle|n\rangle,$ and
$|e_{N}\rangle|n\rangle,$ on the motional excitation number $n,$ which
arises from the spatial variation of the electric field along the $x$
axis \cite{wineland98}. As shown in \cite{gerry97}, if the two laser
beams have a relative phase difference of $\pi$ their amplitudes can
be chosen such that the (spatially) constant terms of the electric
fields associated with the two laser beams cancel each other and the
Rabi frequency $\Omega_{\it dis}^{(n)} \propto \eta_x^2 n/N,$ to
leading order in the Lamb-Dicke parameter \cite{secondsideband}. 
In deriving the pulse transformation we have assumed here the second laser
beam to be phase shifted by $\pi,$ with respect to the first. For a
given $|g_{N}\rangle|n\rangle \Leftrightarrow |e_{N}\rangle|n\rangle$
transition a pulse of duration $t^{(n)}_{\it dis, \pi} =
\pi/\Omega_{\it dis}^{(n)},$ then generates the transformation
\begin{eqnarray}
 U_{\it dis}^{(n)}(t_{0},t^{(n)}_{\it dis, \pi})
	&=& e^{-i \nu_x n t^{(n)}_{\it dis, \pi}}\, 
	|n\rangle\langle n| \nonumber \\[1ex]
	 && \otimes \ \left\{|e_{N}\rangle\langle g_{N}| 
	- |g_{N}\rangle\langle e_{N}|\right\}\,,
 \label{50}
\end{eqnarray}
which is effectively a C-NOT operation between the collective motion
being in either of the states $|0\rangle$ and $|n\rangle,$ and the
electronic state of the $N$th ion. For a motional state $|0\rangle,$
the Rabi frequency $\Omega_{\it dis}^{(0)}=0,$ and the electronic
state {\em remains unaffected\/}. For the motional state $|n\rangle$,
the above transformation flips the electronic excited and ground
state, apart from a phase factor. We note as an important feature of
the C-NOT operation proposed here that it remains valid even beyond the
Lamb-Dicke limit since it's essential feature of not affecting the
motional ground state $|0\rangle$ does not depend on the exact value
of the Lamb-Dicke parameter.

The fourth type of pulse required in our preparation scheme is a
dispersive $\pi$-pulse as described above, but acting on all the ions
simultaneously. The set-up of the two laser beams that generate this
pulse is identical to the single ion case described above but where
the beam waists are assumed sufficiently broad to excite all the ions
with the same strength \cite{turchette98}. We further assume that the
spatial variation of the exciting electric fields along the trap axis
is small compared to the separation of the ions in the trap, i.e.\
$k_x \Delta_x \ll 1,$ where the separation $\Delta_x,$ is typically
several $\mu$m in current trapped ion experiments
\cite{wineland98,hughes97}. This may be realized by having the second
laser beam propagating almost perpendicular to the trap axis or
through a Raman excitation which allows to control the effective wave
vector \cite{turchette98}. Alternatively, one can consider using two
beams whose wave vectors are sensitive to the two (independent) radial
directions and employing the ions' collective motion perpendicular to
the trap axis \cite{privcomm}. In either situation the phase of the
exciting electric fields is equal for all ions \cite{dicke54}, and the
dynamics generated by the pulse is given by the product of the
single-ion time evolution operator given in Eq.\,(\ref{50}). The
dispersive interaction then couples the states
$|J,-J\rangle|n\rangle,$ and $|J,J\rangle|n\rangle,$ and for a given
transition the transformation
\begin{eqnarray}
 U_{\it Ndis}^{(n)}(t_{0},t^{(n)}_{\it Ndis, \pi})
	&=& e^{-i \nu_x n t^{(n)}_{\it Ndis, \pi}}\, 
	|n\rangle\langle n| \nonumber \\[1ex]
 && \otimes \prod_{i=1}^N \left\{|e_{i}\rangle\langle g_{i}| 
	- |g_{i}\rangle\langle e_{i}|\right\}\,,
 \label{60}
\end{eqnarray}
is generated through a pulse duration $t^{(n)}_{\it Ndis, \pi} =
\pi/\Omega_{\it Ndis}^{(n)},$ where $\Omega_{\it Ndis}^{(n)} \propto
\eta_x^2 n/N,$ denotes the Rabi frequency for that transition
\cite{secondsideband}.

We now go through our preparation scheme step by step, starting from
the initial state $|\Psi_{R} (0)\rangle = |J,-J\rangle|0\rangle.$
Since the scheme involves acting on one of the ions separately, we
shall, when necessary write $|J,-J\rangle = |J^\prime,-J^\prime\rangle
|g_{N}\rangle$, where $J^\prime = (N-1)/2.$ First, a resonant
$\pi/2$-pulse is applied to the $N$th ion to produce the state
\begin{equation}
 |\Psi_{R} (t_1)\rangle = \frac{1}{\sqrt{2}} 
	|J^\prime,-J^\prime\rangle
	\left\{|g_{N}\rangle + |e_{N}\rangle\right\}|0\rangle\,,
 \label{70}
\end{equation}
at the time $t_1 = t_{\pi/2}.$ Next, a Jaynes-Cummings $\pi$-pulse
transfers the superposition of the electronic state of the $N$th ion
into the collective motion along the trap axis. From Eq.\,(\ref{40})
with $n=0,$ the resulting state at time $t_2 = t_1 + t^{(0)}_{\it JC,
\pi},$ is
\begin{equation}
 |\Psi_{R} (t_2)\rangle = \frac{1}{\sqrt{2}} |J,-J\rangle\left\{|0\rangle + 
	i\,e^{-i \nu_x t_2} |1\rangle\right\}\,.
 \label{80}
\end{equation}
Then, the superposition of vibrational states is transferred into the
electronic degrees of freedom of {\it all\/} the ions simultaneously
by applying a dispersive $\pi$-pulse to all the ions. As seen from the
pulse transformation $U_{\it Ndis}^{(1)}(t_{2},t^{(1)}_{\it Ndis,
\pi}),$ given in Eq.\,(\ref{60}) this effectuates a C-NOT operation
between the collective vibrational state and the internal states of
all the ions simultaneously, {\it independent\/} of the number of
ions. The state after the pulse is
\begin{equation}
 |\Psi_{R} (t_3)\rangle = \frac{1}{\sqrt{2}} \left\{|J,-J\rangle|0\rangle + 
	i\,e^{-i \nu_x t_3}|J,J\rangle|1\rangle\right\}\,,
 \label{90}
\end{equation}
where $t_3 = t_2 + t^{(1)}_{\it Ndis, \pi}.$ We have now generated the
required superposition between the Dicke states $|J,-J\rangle$ and
$|J,J\rangle.$ The remaining two pulses serve to disentangle the
vibrational and electronic degrees of freedom. The first of those is a
dispersive $\pi$-pulse acting on the $N$th ion. From Eq.\,(\ref{50})
with $n=1,$ the state resulting from this pulse at time $t_4 = t_3 +
t^{(1)}_{\it dis, \pi},$ is given by
\begin{eqnarray}
 |\Psi_{R} (t_4)\rangle &=& \frac{1}{\sqrt{2}} 
	\left\{|J^\prime,-J^\prime\rangle|g_{N}\rangle|0\rangle\right. 
	\nonumber \\
 &&	- \left.i\,e^{-i \nu_x t_4} |J^\prime,J^\prime\rangle
	|g_{N}\rangle|1\rangle\right\}\,.
 \label{100}
\end{eqnarray}
Finally a second Jaynes-Cummings $\pi$-pulse, identical to the one
that led us from Eq.\,(\ref{70}) to Eq.\,(\ref{80}), realizes the
maximally entangled state
\begin{equation}
 |\Psi_{R} (t_5)\rangle = \frac{1}{\sqrt{2}}
	\left\{|J,-J\rangle + |J,J\rangle\right\}|0\rangle\,,
 \label{110}
\end{equation}
at time $t_5 = t_4 + t^{(0)}_{\it JC, \pi},$ leaving the collective
motion along the trap axis in its ground state. In the Schr\"odinger
picture $|\Psi (t_5)\rangle = |\Psi_M\rangle,$ where $\phi = N
\omega_0 t_5,$ in Eq.\,(\ref{20}). The phase $\phi$ in the maximally
entangled state can be controlled by changing the phase of the initial
$\pi/2$-pulse with respect to the other electric fields. Note that the
maximally entangled state is produced deterministically by the
procedure described here. Moreover, all pulses considered here drive
the same $|e\rangle \Leftrightarrow |g\rangle$ transition. This is
important for the experimental realization of our proposal since single
transitions can be made independent of magnetic field fluctuations to
first order \cite{wineland98}.

As we have said earlier, the maximally entangled state generated in
Eq.\,(\ref{110}) may be used in high precision spectroscopy
\cite{bollinger96,huelga97}. In Ref.\cite{bollinger96} Bollinger {\it
et~al.} describe a Ramsey technique where once the maximally entangled
state has been established two Ramsey pulses are applied to all ions
simultaneously and the expectation value of the product operator
$\prod_{i=1}^N \left\{|e_{i}\rangle\langle e_{i}| -
|g_{i}\rangle\langle g_{i}|\right\},$ serves to extract the transition
frequency $\omega_0.$ This is measured by determining the number of
ions in the excited or ground states. In order to not degrade the
signal-to-noise ratio, the uncertainty in this measurement must be
$\ll 1$ atom which requires that the number of ions in the trap be
small. In contrast, the Ramsey technique described by Huelga {\it
et~al.} \cite{huelga97} relies on population measurements on a single
ion and the state generation is an integral part of the modified
Ramsey scheme. In this situation the role of the initial $\pi/2$-pulse
at frequency $\omega_0$ is taken over by a $\pi/2$-Ramsey pulse at
frequency $\omega,$ which is detuned by a small amount $\Delta =
\omega_0 - \omega,$ from the resonance frequency $\omega_0$ which one
aims to determine. We therefore assume that the pulses which prepare
the maximally entangled state are generated by laser beams at the
frequencies $\omega$ and $\omega-\nu_x.$ Thus the frequency $\omega_0$
is replaced by $\omega$ in the pulses, and we describe the time
evolution in a frame rotating with the frequency $\omega$ of the
Ramsey pulses. We denote the state of the system in this frame by
$|\Psi_{R^\prime} (t)\rangle,$ and under the assumption $|\Delta| \ll
\Omega, \Omega_{\it JC}^{(n)}, \Omega_{\it dis}^{(n)}, \Omega_{\it
Ndis}^{(n)},$ the pulse transformations in that frame are the same as
given in Eqs.(\ref{30})-(\ref{60}). Starting from the state
$|\Psi_{R^\prime} (0)\rangle = |J,-J\rangle|0\rangle$ the pulse
sequence described above leads to the state $|\Psi_{R^\prime}
(t_5)\rangle = \left\{ |J,-J\rangle + |J,J\rangle
\right\}|0\rangle/\sqrt{2}.$ We then assume that the system is let
free to evolve for a time $T,$ resulting in the state
$|\Psi_{R^\prime} (t_6)\rangle = \left\{ |J,-J\rangle + e^{-i N \Delta
T} |J,J\rangle \right\}|0\rangle/\sqrt{2},$ where $t_6 = t_5 + T.$
Then the pulse sequence which generates the maximally entangled state
is applied again but in {\it reverse\/} order. This results in the
final state
\begin{eqnarray}
 |\Psi_{R^\prime} (t_7)\rangle &=& \frac{1}{2} |J^\prime,-J^\prime\rangle 
	\left\{ \left(1 +
 	(-1)^N\,e^{-i N \Delta T}\right)|g_{N}\rangle  \right.
	\nonumber \\
 &&	+ \left. \left(1 - (-1)^N\,e^{-i N \Delta T} \right)
 	|e_{N}\rangle \right\}|0\rangle\,,
\end{eqnarray}
at the time $t_7 = T + 2 t_5.$ From this state the resonance frequency
$\omega_0$ can be determined with uncertainty $\delta\omega_0 \propto
1/N,$ \cite{huelga97}, by measuring the internal state of the $N$th
ion, and where $P = \{1 - (-1)^{N}\cos{[N \Delta T]} \}/2,$ gives the
probability of finding the $N$th ion in its excited state.

To address the susceptibility of our scheme to imperfections in the
state generation procedure we emphasize that the experiment reported
in \cite{turchette98} has demonstrated the generation of almost
maximally entangled states through unitary manipulations. The observed
fidelities of approximately $0.7$ indicate that deviations from
unitary evolution do not destroy the sought-after state and affirm
the validity of the unitary analysis presented here. 

In summary, we have proposed an efficient method for generating
maximally entangled internal states of a system of $N$ trapped
ions. The method has the further advantage of being
deterministic. Finally, we have shown how a such a state and the
generation scheme described here may be used for high precision Ramsey
spectroscopy.

\section*{Acknowledgements}

This work was supported in part by the UK Engineering and Physical
Sciences Research Council and the European Community. C.C.G.\
acknowledges the support of a PSC-CUNY grant and a Schuster Award from
Lehman College. J.S.\ is supported by the German Academic Exchange
Service (DAAD-Doktorandenstipendium aus Mitteln des dritten
Hochschulsonderprogramms). We wish to thank Drs.\ S.F.~Huelga, M.B.~Plenio,
P.L.~Knight and D.J.~Wineland for helpful discussions.

\end{multicols}


\begin{thebibliography}{99}

 \bibitem{quantcomp} A.~Barenco, Contemp.\ Phys.\ {\bf 37}, 375
 (1996); R.J.\ Hughes {\it et~al.}, {\it ibid}, {\bf 36}, 149 (1995);
 M.B.~Plenio and V.~Vedral, {\it ibid}, to be published (1998).

 \bibitem{freqstd} D.J.~Wineland {\it et~al.}, Phys.\ Rev.\ A {\bf
 46}, R6797 (1992); W.M.\ Itano {\it et~al.}, {\it ibid}, {\bf 47},
 3554 (1993); D.J.~Wineland {\it et al.}, {\it ibid}, {\bf 50}, 67
 (1994); D.J.~Wineland {\it et al.}, e-print quant-ph/9809028.

 \bibitem{wineland98} D.J.~Wineland {\it et al.}, Journal of Research 
 of NIST, {\bf 103}, 259 (1998).

 \bibitem{bollinger96} J.J.~Bollinger {\it et al.}, Phys.\ Rev.\ A,
 {\bf 54}, R4649 (1996).

 \bibitem{spec} D.J.~Wineland {\it et al.}, IEEE Transactions on
 Ultrasonics, Ferroelectrics and Frequency Control, {\bf 37}, 515
 (1990). D.J.~Berkeland {\it et al.}, Phys.\ Rev.\ Lett.\ {\bf 80},
 2089 (1998).

 \bibitem{dicke54} R.H.~Dicke, Phys.\ Rev.\ {\bf 93}, 99 (1954).

 \bibitem{GHZ} D.M.~Greenberger, M.A.~Horne, and A.~Zeilinger, in {\it
 Bell's Theorem, Quantum Theory, and Conceptions of the Universe,\/}
 edited by M.~Kafatos (Kluwer, Dordrecht, 1989); D.M.~Greenberger,
 M.A.~Horne, A.~Shimony, and A.~Zeilinger, Am.\ J.\ Phys.\ {\bf 58},
 1131 (1990).

 \bibitem{mermin90} N.D.~Mermin, Phys.\ Rev.\ Lett.\ {\bf 65}, 1838
 (1990).

 \bibitem{atomcat} C.C.~Gerry and R.~Grobe, Phys.\ Rev.\ A {\bf 56},
 2390 (1997); {\it ibid}, {\bf 57}, 2247 (1998); G.S.~Agarwal,
 R.R.~Puri, and R.P.~Singh, {\it ibid}, {\bf 56}, 2249 (1997);
 M.G.~Benedict, {\it et~al.}, Acta Phys.\ Slovaka {\bf 47}, 259
 (1997).

 \bibitem{turchette98} Q.A.~Turchette {\it et al.}, Phys.\ Rev.\ Lett.\ 
 {\bf 81}, 3631 (1998).

 \bibitem{huelga97} S.F.~Huelga {\it et al.}, Phys.\ Rev.\ Lett.\ 
 {\bf 79}, 3865 (1997).

 \bibitem{ramsey63} N.F.~Ramsey, {\it Molecular Beams,} (Oxford
 University, London, 1963).

 \bibitem{cirac95} J.I.~Cirac and P.~Zoller, Phys.\ Rev.\ Lett.\ {\bf
 74}, 4091 (1995).

 \bibitem{steinbach97} J.~Steinbach, J.~Twamley, and P.L.~Knight,
 Phys.\ Rev.\ A {\bf 56}, 4815 (1997).

 \bibitem{deterministic} See \cite{turchette98} and references
 therein.

 \bibitem{hughes97} R.J.~Hughes {\it et al.}, 
 Fortschritte der Physik {\bf 46}, 329 (1998). For the
 two particle case, Turchette {\it et~al.}  \cite{turchette98} have
 experimentally demonstrated an alternative technique to
 differentially address single ions by controlling the ion
 micromotion.

 \bibitem{king98} B.E.~King {\it et al.}, Phys.\ Rev.\ Lett.\
 {\bf 81}, 1525 (1998).

 \bibitem{gerry97} C.C.~Gerry, Phys.\ Rev.\ A {\bf 55}, 2478 (1997).

 \bibitem{knight86} P.L.~Knight, Phys.\ Scr.\ T {\bf 12}, 51 (1986);
 S.J.D.~Phoenix and P.L.~Knight, J.\ Opt.\ Soc.\ Am B {\bf 1}, 116
 (1990).

 \bibitem{harmonic} We neglect the ions' micromotion whose main effect
 is the alteration of transition rates between quantum levels
 \cite{bardroff96} which can be accounted for by experimental
 calibration \cite{wineland98}.

 \bibitem{bardroff96} P.J.~Bardroff {\it et al.}, Phys.~Rev.~Lett.\ 
 {\bf 77}, 2198 (1996).
 
 \bibitem{LDL} W.E.~Lamb, Phys.~Rev.\ {\bf 51}, 187 (1937);
 R.H.~Dicke, Phys.~Rev.\ {\bf 89}, 472 (1953).
 
 \bibitem{cirac96} J.I.~Cirac {\it et al.}, 
 Adv.\ At.\ Mol.\ Opt.\ Phys.\ {\bf 37}, 237 (1996).

 \bibitem{secondsideband} The Rabi frequency here is reduced through
 its dependence on $\eta_x^2/N,$ which may be compensated for by high
 enough laser power. Second sideband excitation has been demonstrated
 for a single ion in Ref.\cite{meekhof96}.

 \bibitem{meekhof96} D.M.~Meekhof {\it et al.}, Phys.\ Rev.\ Lett.\
 {\bf 76},1796 (1996).

 \bibitem{privcomm} C.R.~Monroe and D.J.~Wineland, private
 communication.

\end{thebibliography}
\end{document}